\documentclass[aps,prb,preprint]{revtex4-1}
\usepackage{dcolumn}%
\usepackage{bm}%
\usepackage{graphicx}
\usepackage{amsfonts}
\usepackage{amsmath}
\usepackage{amssymb}
\usepackage{hyperref}
\usepackage{indentfirst}
\usepackage[usenames,dvipsnames]{pstricks}
\usepackage{epsfig}
\usepackage{pst-grad} 
\usepackage{pst-plot} 

\begin{document}

\title{Using river locks to teach hydrodynamic concepts}
\author{Vagson L Carvalho-Santos}
\author{Thales C Mendes}
\author{Enisvaldo C Silva}
\author{M\'arcio L Rios}
\author{Anderson A P Silva}
\affiliation{Instituto Federal de Educa\c c\~ao, Ci\^encia e Tecnologia Baiano - Campus Senhor do Bonfim, Bahia, Brazil}
\begin{abstract}In this work, it is proposed the use of a river lock as a non-formal setting for teaching hydrodynamical concepts. In particular, we describe the operation of a river lock situated in the Sobradinho's dam, at the S\~ao Francisco River (Brazil). A model to represent and to analyse the dynamics of a river lock operation is presented and  we derive the dynamical equations for the rising of the water column as an example to understand the Euler's equation. Furthermore, with this activity, we enable an integration between the content initially introduced in the classroom with practical applications, allowing the association of physical themes to contents relevant in other disciplines such as History and Geography. Besides this, experiences of this kind enable teachers to talk about the environmental and social impacts caused from the construction of a dam and, consequently, crossover of concepts have been possible, making learning more meaningful for the students.
 \end{abstract}

\maketitle


\section{Introduction}
According to the National Curricular Parameters (\textit{Par\^ametros Curriculares Nacionais}) \cite{PCN} of Brazil, the student's formation must have, as the main objective, the getting and comprehension of basic concepts, as well as the scientific preparation and the capacity to use different technologies related to the different actuation areas. In the secondary level, it is proposed a most general formation, opposite to that specific one. It is also proposed the development of the capacity to research, analyse and select informations. Furthermore, education must aim develop the capacity to learn, create and formulate, instead of memorization exercise.

In this way, one can not restrict education to the scholar context, once it can happen in several places, from which we can cite scientific disseminate centers and museums \cite{Tran-Tesis,Ash-Book}. Moreover, it is known that the student's knowledge is gotten not only in experiences occurring inside the classroom, but also from their experiences in the everyday life and didactic activities, which can be proposed, for example, by the teacher. Thus, the scientific dissemination centers, the electronic media and the science museums are important tools in the learning process \cite{museuselet}. According to Langhi \textit{et al} \cite{langhi}, learning may happen in several environments, and these ones can be classified as formal, informal and non-formal settings. Furthermore, we can cite, as learning activity, the activities known as scientific popularization.

Langui \textit{et al} \cite{langhi} define formal education as that one occurring in the scholar environment and others teaching establishments, with own structure and planning, whose objective is to work, didactically, the systematized knowledge. The non-formal education is that one having collective character and involving educative practicals which does not occur in the scholar environment and has not legal requirement. During these activities, the student experiments the freedom to choose the concepts to be learned. Among the examples offering non-formal settings, one can cite: museums, media, training agencies for specific social groups and non-conventional teaching institutions, which organize events such as free courses, science fairs and scientific meetings \cite{langhi}. It is important to say that, in spite it does not occur in a scholar environment, the non-formal education is not free of a certain degree of intentionality and systematization.

In this context, museums and science centers, which can be classified as non-formal education settings \cite{langhi}, can favour the conceptual expansion and refinement in an environment which can bring emotions, becoming coupled with the cognitive process, doted of an intrinsic motivation for the learning of science \cite{emo8}. Furthermore, non-formal settings are places providing the appreciation and the understanding of sciences by voluntary and individual actions, popularizing the scientific and technological knowledges \cite{cient9}. And still, nowadays, it is increasing the concern with both, affective and emotional impacts and with the production of meaning and knowledge construction \cite{marandino}.

Given the importance of non-formal settings for the student's learning, there are several works dedicated to research the influence of museums for the science education (see, for example Refs. \cite{Ash-Book,Tran-Tesis,cient9,solinf,park}). Nevertheless, the presence of museums and science centers are not a reality in the most of Northeast of Brazil. Thus, to visit one of these non-formal spaces must be a tiring, stressful and expensive activity, due the distances among several cities located in this region and the nearest scientific centers. In this case, in order to provide different learning environments for the students, the teacher can propose accessible activities in his living region and to arouse the enthusiasm of the student for scientific knowledge. 

When we think in hydrodynamics teaching, several works have investigated about the learning process and new forms to address concepts on this issue. For example, one can cite some elaborated experiments destined to teach fluid dynamics \cite{turblam,airjets}. Furthermore, Arellano \textit{et al} have proposed the injection of particles or bubbles in order to simplify the complex task of teaching the hydrodynamics of a swimmer’s propulsion to undergraduate students of Physical Education, once, with this technique, the students have the opportunity to see how the water is actively moving when the body is propelled through the water \cite{Swimming}. In addition, from the analysis of 15 original simulations, created with GeoGebra software, Romero \textit{et al} \cite{romero-EJP} have applied a questionnaire on the interest of using simulations to teach fluid mechanics to simulation-taught students and compare the answers to that given by students taught without use of simulations. At the examination, the average grade and the percentage of passed students were higher in group 1 than in group 2, however, the author recognises that additional strategies need to be adopted aiming to help students develop the skills required to succeed in physics course. An apparent paradox in communicating vessels systems is discussed in the work of Miranda \cite{Miranda}, in which the author shows that, for a liquid in any connected vessel system, it is not possible to realize simultaneously Pascal's principle, mass and energy conservation. In addition, there are propositions of hydrostatic teaching from experimental activities using low cost materials, e.g., plastic bottles \cite{Pet} and water cup \cite{dinner}. Finally, the transport of water from the roots to the crown of trees is discussed for two conduit architectures is considered in the work of \cite{TreeHyd} and it is proposed to be exposed to undergraduate students, in order to get an interdisciplinar communication with Biology. In this work, the author considers the subject of broad interest because it provides a naturally-occurring example of an unusual metastable state of matter.

In this Paper, we report the proposition to study basic concepts of hydrodynamics from the visit to a river lock, which consists in a work of hydraulic engineering that allows boats ascend or descend the rivers and seas in places where there are gaps (dam or waterfalls). In our case, we have chosen to visit the Sobradinho's river lock, once it is the nearest structure of this kind to Senhor do Bonfim - BA, Brazil. Furthermore, we show the proposition to construct a model of a river lock, in order to study the operation of this system, in more details. We also discuss the opportunity that must be given to the students, to discuss and couple knowledge of different contents of other disciplines, such as Geography and History, to Physics, seeing the science as a human activity and understanding about social, economic and politic impacts brought from the construction of a dam.

This work is divided as follows: in section 2, we present the Sobradinho's river lock and talk about its operation and economic importance; section 3 brings a discussion on some hydrodynamic concepts that can be addressed from the operation of a river lock; in section 4, we present the proposition to construct a model of a river lock in order to study, in more details, the river lock's operation; finally, in section 5, the conclusions are presented.

\section{The Sobradinho's dam and hydrodynamic of river locks}
The artificial lake formed from the Sobradinho's dam has length of 320 km \cite{ashfra} (from municipality of Sobradinho to the municipality of Pilão Arcado) and a water surface area of 4,214 km$^2$. Its storage capacity is around 34.1 million of liters, being the second largest artificial lake in the world. It ensures, through a depletion of up to 12 m, together with the Tr\^es Marias Reservoir, a regulated flow of 2,060 m$^3$/s during the dry season, allowing operation of all hydroelectric plants of the \textit{Companhia Hidroel\'etrica do Vale do S\~ao Francisco} (CHESF), situated along the S\~ao Francisco River. 

\begin{figure}[h]
\begin{center}
\includegraphics[scale=0.9]{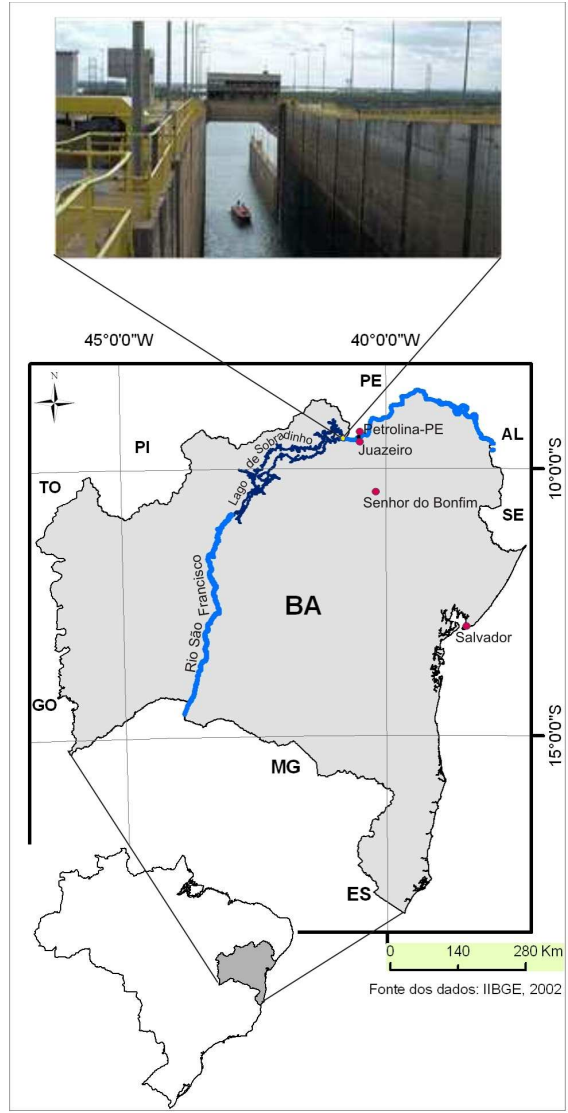}\end{center}
\caption{Location of the Sobradinho's dam. Featured, we have a view of the river lock in that dam. The vessel is raised to the highest level by opening a gate, which releases the entry of water coming from the dam in the region of the lock.}\label{sobr}
\end{figure}

This dam incorporates a river lock, owned by \textit{Companhia Docas do Estado da Bahia} (CODEBA), whose camera has 120 m of length and 17 m width, allowing the boats to overcome the gap created by the dam, around 32.5 meters, with maximum filling time of 16 minutes. This river lock ensures the continuity of traditional navigation between the stretch of the São Francisco River between the cities of Pirapora-MG and Juazeiro-BA/Petrolina-PE (1,371 km navigable) favouring the waterway transport and therefore, the commercial navigation in the Old Chico (form with which the river is known in the region). In Figure \ref{sobr}, one can see a highlighted view of the river lock under discussion. In this same figure, we show a map locating the Sobradinho's dam in the state of Bahia-Brazil.

\begin{figure}[h]\begin{center}
\includegraphics[scale=0.4]{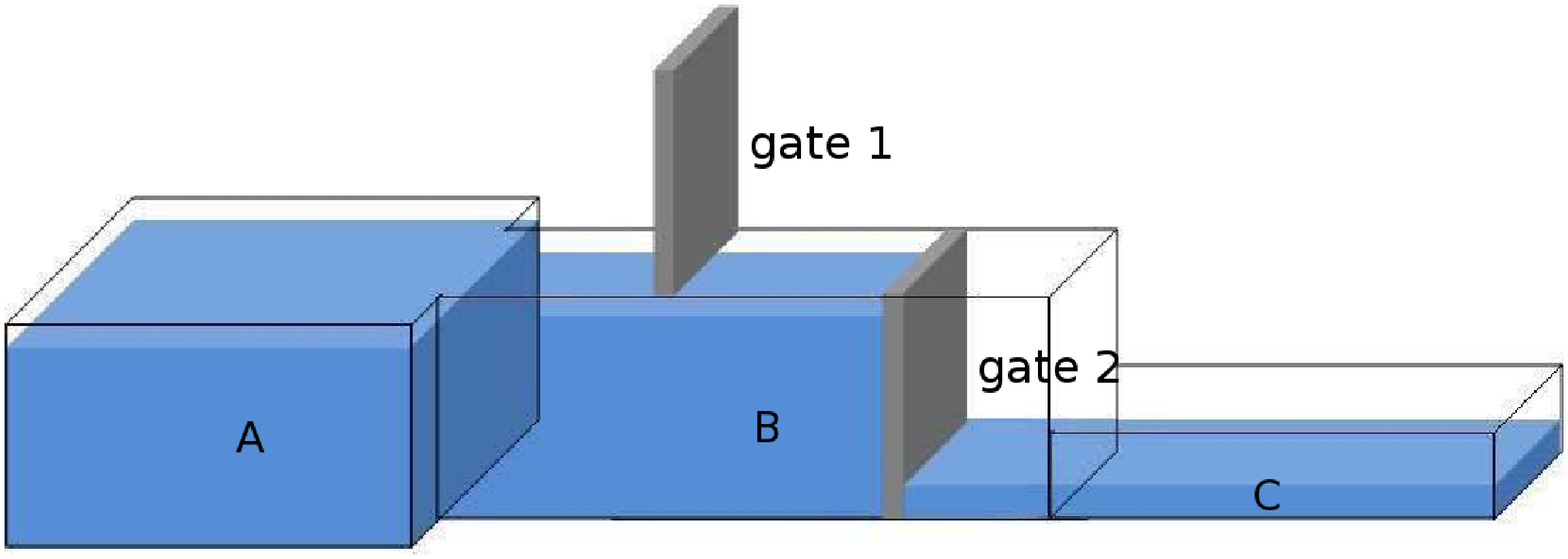}\end{center}\caption{Schematic view of the operation of a river lock. Once, in statical conditions, the pressure must be the same for points situated in the same height, if we open the link between A and B, there will be a water flow from the region A to the region B, until the water level is equal in the two sides. The same argument must be used when we open the link between B and C. In this case, when the board comes from C, going to A, the link between B and C must be opened in order to decreases the water level in B. The gate is opened and the boat can go to B. Now, the link between A and B is opened and the water level in B rises. The gate 1 is opened and the boat can pass to A.}\label{pontos}
\end{figure}

River locks function as stairs or elevators for ships or boats, in which there are two gates separating the two river (or sea) levels. In the Fig. \ref{pontos}, we show a schematic view representing, in a very simplified form, the operation of a river lock. When the boat ups the river, it enters in the lock at the downstream side (marked in the figure as C) and remains in the chamber (region B). The downstream is then closed and the chamber filled with water, causing the boat to rise until it reaches the level of the upper reservoir. Thereafter, the gate 1 can be opened and the boat leaves the lock, going to the dam, marked in the Figure as the region A. When the boat downs the river, it enters in the chamber at the upstream side of the lock and the gate is closed, emptying the chamber gradually until it reaches the level of the lower reservoir. Finally, the gate 2 is opened and the boat leaves of the river lock. The operations of filling and emptying the chamber are usually made by gravity with the help of small gates and valves. An animation showing the operation of the Sobradinho's river lock can be found on the website cited in Ref. \cite{ashfra}.

\subsection{Static}
In order to study the hydrodynamical principle of the operation of a river lock, we will simplify it to a communicating vessels system, whose dynamical properties can be well defined from the Euler equation \cite{landau,faber}:
\begin{equation}
\frac{\partial \mathbf{v}}{\partial t}+(\mathbf{v}\cdot\nabla)\mathbf{v}=-\frac{\nabla p}{\rho}+g\mathbf{\hat{z}},
\end{equation}
where $\mathbf{v}$ is the fluid velocity, $\rho$ is its density, $p$ the pressure of the fluid and $g$, the gravity acceleration. For a fluid in rest, we have:
\begin{equation}\label{static}
\nabla p=\rho g\mathbf{\hat{z}}.
\end{equation}
If the fluid density is considered constant along its volume and the $z$ axis is taken as vertical, the Eq. (\ref{static}) can be integrated to give:
\begin{equation}
\frac{\partial p}{\partial x}=\frac{\partial p}{\partial y}=0,\hspace{1cm}\frac{\partial p}{\partial z}=-\rho g.
\end{equation}
Thereby,
\begin{equation}\label{hidro}
p=-\rho g z + \text{constante}.
\end{equation}
If the fluid has a free surface, in the height $h$, for which an external pressure $p_0$, at all points, is applied, this surface must be a horizontal plane $z=h$. From the condition $p=p_0$ when $z=h$, we have that the constant, in the Eq. (\ref{hidro}), is given by $p_0+\rho g h$, such that:
\begin{equation}\label{equil}
p=p_0+\rho g (h-z).
\end{equation}

In this way, given a point in a fluid, the pressure on this point will depend only on the height of the liquid. In this case, from observing the Fig. \ref{pontos}, we can conclude that, from the opening of the gate 2, a point situated in the region C will be subject, initially, to a pressure lower than the pressure on a point, at the same height, situated in the region B. From the Eq. (\ref{equil}), if the regions are linked, the pressure must be equal in the two points, thus, there will be a transference of fluid from the region B to C, when connecting valve between these two regions is opened. The region C does not rise its level because the water goes to the river, and, in the case of Sobradinho's dam, it goes to the cities of Petrolina/Juazeiro. Then, there is a reduction on the water height in the region B, until it reaches the level of the river (region C).

From the same principle, when the region B is in the level of the river (C), there will be a flux of water from the region A to the region B, filling this region and rising the water height.

\subsection{Dynamics}
Now, we will analyse the rising velocity of the liquid column inside the river lock in function of the water flow released by the connection valve between the regions A and B. Obviously, if we maintain a constant flow $\mathcal{Q}$, given by $\mathcal{Q}=\frac{d V}{dt}$, where $V$ is the volume of water that pass from the region A to the region B, a boat will be rise until the upper of the river lock with a constant velocity, given by:
\begin{equation}
v(t)=\frac{\mathcal{Q}}{S_B},
\end{equation}
where $S_B$ is the surface's area of the region B. In this case, the height will be a linear function of the time, given by:
\begin{equation}\label{MRU}
z(t)=\frac{\mathcal{Q}}{S_B}t,
\end{equation}
where we have done $z(0)=0$. 

\begin{figure}[h]
\begin{center}
\includegraphics[scale=0.4]{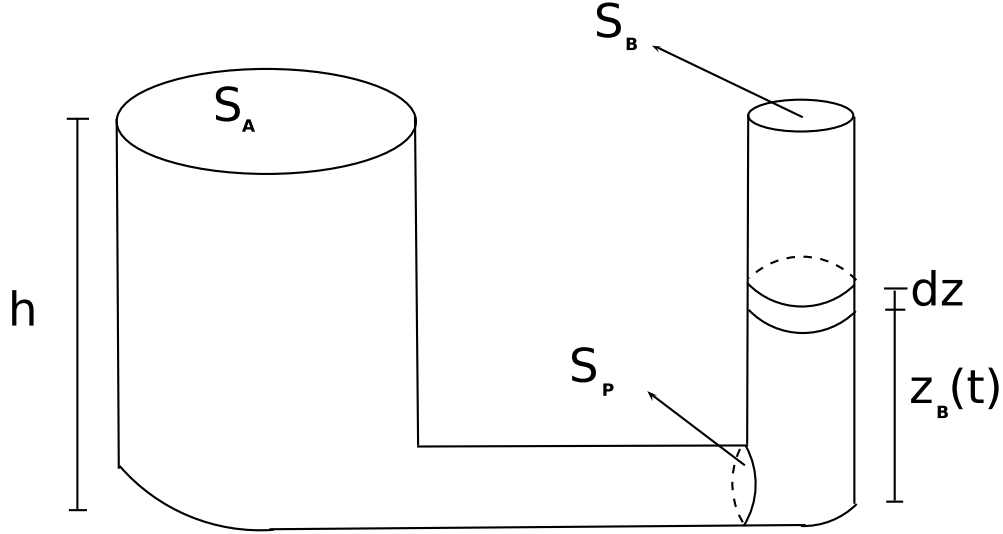}\end{center}
\caption{Simple model to analyse the rising velocity of the liquid column in a river lock. Here, we have considered that the water flow is a function of the height $z$, inside the region B represented in the Fig. \ref{pontos}. Once the pressure in the opening linking the regions A and B increases with $z$, the velocity of exit of water in this opening diminishes, thus, the liquid column rises with a decreasing velocity.}\label{novesq}
\end{figure}

Now, we will study what occurs when the water flow is variable with the time, which can be obtained, for example, by maintaining the opening of the valve linking A and B with a constant area $S_P$, as it is shown in Fig. \ref{novesq}. In this case, when the valve is opened, the water level in the region B rises, increasing the pressure of the water on the exit $S_P$. Thereby, if we maintain the valve opened with a constant area, the water flow will not be constant, once the increasing of pressure in the region B will diminish the water flow through the opening that separates the two regions. 

As it has been said, when the link between A and B is opened, there will be a water flow from the region A, with area $S_A$, to the region B, with area $S_B$, passing through  the opening with area $S_P$. In order to determine the velocity of the water when it passes through the opening, we will use the Bernoulli's equation \cite{landau}:
\begin{equation}
\frac{1}{2}\rho v^2+\rho gz+p=\kappa,
\end{equation}
where $\kappa$ is a constant. In this case, the fluid's velocity through the opening with area $S_P$ is given by:
\begin{equation}\label{continuidade}
\frac{1}{2}\rho v_P^2+\rho gz_P+p_P=\frac{1}{2}\rho v_S^2+\rho gz_A+p_A,
\end{equation}
in which the subscript indices $P$ and $A$ are representing the surfaces with area $S_P$ and $S_A$, respectively. $z_{P}\equiv z(t)$ is the height of water column when it pass to the region with area $S_B$. From the Fig. \ref{novesq}, we have that $z_A=h$, $z_P=0$, $p_A=p_0$, $p_P=p_0+\rho g z_B(t)$. Furthermore, we will consider $S_A\gg S_P$, then the lowering velocity of the water in the region A is $v_A=0$. Thus, from Eq. (\ref{continuidade}), the velocity of the water through the valve linking the regions A and B will be given by:
\begin{equation}
v_P(t)=\sqrt{2g[h-z_B(t)]}.
\end{equation}
In this way, defining $z_B(t)\equiv z(t)$, we have that the water flow $\mathcal{Q}(t)$ through the region with area $S_P$ is:
\begin{equation}
\mathcal{Q}(t)=S_P\sqrt{2g[h-z(t)]}.
\end{equation}
And, in this case, the rising velocity of the water at the river lock (region B) will be:
\begin{equation}\label{vel}
v_B(t)=\frac{\mathcal{Q}(t)}{S_B}=\frac{S_P}{S_B}\sqrt{2g[h-z(t)]}.
\end{equation}
Once $z(0)=0$, we have that the velocities of the water through the valve, $v_P(t)$ and in the river lock region, $v_B(t)$, in the instant $t=0$ are given by $v_P(0)=\sqrt{2gh}$ and $v_B(0)=\frac{S_P}{S_B}\sqrt{2gh}$. Furthermore, they decrease their values when the height of the water column rises, in such way that these velocities will vanish when $z(t)=h$.

Finally, aiming to determine the function, $z(t)$, for which the height of the water rises in the region B, we start from the velocity definition $v\equiv\frac{dz}{dt}$. One can note that:
\begin{equation}
\int\frac{dz}{\sqrt{2g[h-z(t)]}}=\frac{S_P}{S_B}t.
\end{equation}
This integral is evaluated to give: 
\begin{equation}
z(t)=h+\left(\kappa\frac{S_P}{S_B}\right)t-\frac{1}{2}g\left(\frac{S_P}{S_B}\right)^2t^2 -\frac{\kappa^2}{2g},
\end{equation}
where $\kappa$ is a constant of integration. Taking the initial boundary condition $z(0)=0$, we obtain $\kappa=\sqrt{2gh}$, and so:
\begin{equation}\label{MRUV}
z(t)=\sqrt{2gh}\left(\frac{S_P}{S_B}\right)t-\frac{1}{2}g\left(\frac{S_P}{S_B}\right)^2t^2.
\end{equation}

One can note that the rising of the water level in the river lock is well represented by a uniformly variable rectilinear motion function, in which the water, and consequently the boat, begin its upward movement with initial velocity $v_B(0)=\frac{S_P}{S_B}\sqrt{2gh}$, decreasing its value with a constant acceleration given by $a=\left(\frac{S_P}{S_B}\right)^2 g$. In this way, the needed time to the boat reaches the highest point of the river lock, $z(t)=h$, and continue its flux, is:
\begin{equation}
t_\text{up}=\frac{S_B}{S_P}\sqrt{\frac{2h}{g}}.
\end{equation}
As expected, the rising time of the boat is directly proportional to the river lock surface area, $S_B$, and decreases with the increasing of the height of the river lock.

\section{Constructing a river lock's model}
The visit to the river lock has been important because, in this ambient, the students can have contact with several devices that could not be viewed in a formal space (classroom). Furthermore, they have the opportunity to hear and discuss the theoretical and practical aspects about the operation of a river lock with a professional having experience with the possible problems coming from the operation of these structures. Other advantage of this visit comes from the opportunity to observe probable social and environmental impacts caused by the construction of a dam.

\begin{figure}[h]\begin{center}
\includegraphics[scale=0.8]{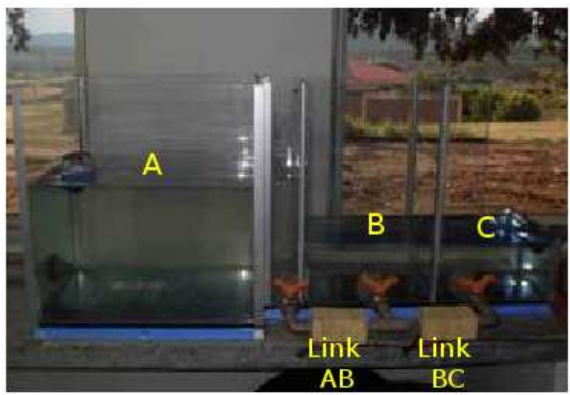}\end{center}\caption{Photograph of the model constructed in order to observe, inside the classroom, the operation of a river lock. Here, we highlight the regions A, B and C, represented in the schema of the Fig. \ref{novesq}.}\label{fotomaq}
\end{figure}

After the visit to the Sobradinho's dam, as a part of a multidisciplinary activity involving several areas of knowledge, as Physics, History, Geography and Portuguese, we have constructed a model of a river lock (See Fig. \ref{fotomaq}), in order to evaluate the learning process and systematize the acquired knowledge. The model was constructed according the schema of the Fig. \ref{pontos}. In order to avoid water waste, when it pass from the region B to the region C, we have put a vessel receiving the water exiting from the model. A water pump returned the water from the vessel to the region A. 

In the Fig. \ref{fotomaq}, we present a photograph of the constructed model. In that, one can note highlighted, the regions A, B and C. The valves linking the regions has been done through pipes and taps (Link AB and Link BC), which release water passage, when opened. When the link AB is opened, we release the water flow from region A to the region B. In this case, the liquid column height in B rises. In this model, in order to ensure the constant height of the region A, we have filled the vessel while the water have flowed to the region B. This procedure will ensure that the rising of the water in the region B obeys the Eq. (\ref{MRUV}), that is, the liquid column is rising in an uniformly variable rectilinear motion. Obviously, by closing the link AB and opening the link BC, the water flows from B to C, however, once the height at the region C does not rise, the passage of the water from B to C must obey the Eq. (\ref{MRU}), once the flow must be considered constant.

Aiming to test the validity of the Eq. \ref{MRUV}, we performed some measurements with the constructed model. The model's parameters are $h=34.5$ cm, $S_P=\pi$ cm$^2$ and $S_B=847.9$ cm$^2$. To predict the needed time to fill the place B, we used $g=9.81$ m/s$^2$, obtaining $t_{\text{up}}\approx 71.58$ s. Besides to measure the filling time, we determined the height of the water column for $t=30$ s and $t=60$ s. The predicted values are $z(30\text{ s})\approx22.86$ cm and $z(60\text{ s})\approx33.60$ cm. The obtained experimental values were obtained after ten measurements an they were $z(30\text{ s})=21.5\pm1.4$ cm, $z(60\text{ s})=31.8\pm1.3$ cm and the filling time was $t_{up}=69.72\pm0.7$ s. The disagree among the theoretical and experimental values is associated with errors due to sealing problems in the gates, which made ​​it possible that there was a water flow slightly higher than expected from the A to B region. In addition, there may be variations due to the actual value for the local gravity acceleration.

\section{Conclusions}
We have proposed the visit to a river lock as a non-formal setting for teaching hydrodynamics. In particular, we have visited the Sobradinho's dam, situated in the S\~ao Francisco river, 150 km away from Senhor do Bonfim. This travel had, as main objective, to show the operation of a hydroelectric plant and, in special, a river lock.

With this activity, we believe that it can be enabled the integration among the propaedeutic contents, learned in the classroom, and practical applications of these ones. In the local, we have realized a discussion on the physical principles based on the operation of a river lock, as well as its economic importance, in such way that students might realize that physics is not a subject disconnected from reality, having important links with other areas of knowledge. Thus, the students realized that the economic development is linked to the scientific development of a country. When we return to the formal space (classroom), we have constructed a model to show, in a more detailed form, the operation of a river lock for the students.

The visit to the river lock, with the participation of students and teachers, allowed a greater integration among stakeholders and the contextualization of contents relevant to each discipline. As a consequence, it was accomplished with a interdisciplinarity among Physics, History and Geography, where it has been possible to discuss on problems such as siltation, transportation, economy, water use for irrigation and electricity generation, as well as social, economical and environmental impacts due the construction of a dam. Furthermore, the History teacher could explain the evolution of the economy and culture of the communities located in places affected by the Sobradinho's dam. 

Finally, besides the river lock, the presence of a hydroelectric plant at Sobradinho gave us the opportunity to discuss other physical concepts with the students, e.g., the processes of electric power generation, hydropower, alternative forms for energy generation and electromagnetic induction. We could review the energy conservation principle. 

\section*{Acknowledgements}
The authors thank CNPq (grant number 562867/2010-4) and PROPES of IFBaiano for financial support. We thank also the employees of the Sobradinho's river lock for the receptivity and hospitality. We are also in debit with Antonio S Silva, Maisa F S Martins and Amanda A Melo for helping with the river lock model operation.


\thebibliography{99}



\bibitem{PCN}
http://portal.mec.gov.br/seb/arquivos/pdf/CienciasNatureza.pdf (In Portuguese).

\bibitem{Tran-Tesis}
Tran L U 2002 \textit{The roles and goals of educators teaching science in non-formal settings}, thesis submitted to the Graduate Faculty of North Carolina State University. Available in http://repository.lib.ncsu.edu/ir/bitstream/1840.16/830/1/etd.pdf. Accessed in May 13, 2013.

\bibitem{Ash-Book}
Ash D and Wells G 2006 \textit{Dialogic Inquiry in Classroom and Museum}, In \textit{Learning in places: the informal education reader} edited by Zvi B, Burbules N C,  Silberman-Keller D 2006, p. 35. Peter Lang Publishing Inc. New York.

\bibitem{museuselet}
Falk J H 2001 {\it Free-Choice Science Learning: Framing the discussion} (New York: Teachers College Press).

\bibitem{langhi}
Langhi R and Nardi R 2009 \textit{Rev Bras Ens F\'is} {\bf 31} 4402 (In Portuguese).

\bibitem{emo8}
Queiroz G, Krapas S, Valente M E, David E, Damas E and Freire F 2002 \textit{Rev Bras Pesq  Ens Ci\^enc} {\bf 2} 2 (In Portuguese)

\bibitem{cient9}
Pereira G R and Coutinho-Silva R 2010 \textit{Rev Bras Ens F\'is} {\bf 32} 3402 (In Portuguese)

\bibitem{marandino}
Marandino M 2006 {\it A Pesquisa em Ensino de Ci\^encias no Brasil e suas Metodologias} (Editora da Uniju\'i, ́Iju\'i) (In Portuguese)

\bibitem{solinf}
Aroca S C and Silva C C 2011 \textit{Rev Bras Ens F\'is} {\bf 33} 1402 (In Portuguese)

\bibitem{park}
Moll R F 2010 \textit{Phys Educ} {\bf 45}, 362

\bibitem{turblam}
Riveros H G and Riveros-Rosas D 2010 \textit{Phys Educ} {\bf 45} 288

\bibitem{airjets}
L\'opez-Arias T, Gratton L M, Zendri G and Oss S 2011 \textit{Phys Educ} {\bf 46} 373

\bibitem{Swimming}
Arellano R and Pardillo S 2001 \textit{Teaching Hydrodynamic Concepts Related to Swimming Propulsion Using Flow Visualization Techniques in the Pool}. Available in http://www.ugr.es/~swimsci/SwimmingScience/page4/page16/page38/files/2001ArellanoNSTBS.pdf
Accessed in April 10, 2013.

\bibitem{romero-EJP}
Romero C and Mart\'inez E 2013 \textit{Eur. Journ Phys.} \textbf{34} 873

\bibitem{Miranda}
Miranda E N 2009 \textit{Eur J Phys} \textbf{30} L55

\bibitem{Pet}
Jesus V B L and Macedo-Junior M A V 2011 \textit{Rev Bras Ens F\'is} {\bf 33} 1507 (In Portuguese)

\bibitem{dinner}
Marshall R 2013 \textit{Phys Educ} \textbf{48} 390

\bibitem{TreeHyd}
Denny M 2012 \textit{Eur J Phys} \textbf{33} 43

\bibitem{ashfra}
AHSFRA 2011 {\it Administra\c c\~ao da Hidrovia do S\~ao Francisco}. Available in \underline{www.ahsfra.gov.br} (In Portuguese) Acessed in March 13, 2013




\bibitem{landau}
Landau L D and Lifshitz E M 1986 \textit{Fluid Mechanics}, 2$^{nd}$ edition, (Pergamon Books Ltd, Moscow)

\bibitem{faber}
Faber T E 1995 \textit{Fluid Dynamics for Physicists} (Cambridge University Press, Cambridge)
\end{document}